\documentclass[usegraphicx,usenatbib,useAMS]{mn2e}
\usepackage{times}
\usepackage{amssymb}
\begin{document}

\title[Baryonic Conversion Tree]{Baryonic Conversion Tree: The global assembly of stars and dark matter in galaxies from the SDSS}
\author[Jimenez et al.]
{Raul Jimenez$^1$, Benjamin Panter$^2$, Alan F. Heavens$^2$, Licia Verde$^1$ \\
$^1$Dept. of Physics and Astronomy, University of Pennsylvania, Philadelphia,
PA 19104, USA.\\
$^2$Institute for Astronomy, University of Edinburgh, Royal
Observatory, Blackford Hill, Edinburgh EH9-3HJ, UK.}

\maketitle

\begin{abstract}
  Using the spectroscopic sample of the SDSS DR1 we measure how gas
  was transformed into stars as a function of time and stellar mass:
  the baryonic conversion tree (BCT). There is a clear correlation
  between early star formation activity and present-day stellar mass:
  the more massive galaxies have formed about 80\% of their stars at
  $z>1$, while for the less massive ones the value is only about 20\%.
  By comparing the BCT to the dark matter merger tree, we find
  indications that star formation efficiency at $z>1$ had to be about
  a factor of two higher than today ($\sim 10\%$) in galaxies with
  present-day stellar mass larger than $2 \times 10^{11}M_\odot$, if
  this early star formation occurred in the main progenitor.
  Therefore, the LCDM paradigm can accommodate a large number of red
  objects.
  On the other hand, in galaxies
  with present-day stellar mass less than $10^{11}$ M$_{\odot}$,
  efficient star formation seems to have been triggered at $z \sim
  0.2$. We show that there is a characteristic mass (M$_* \sim
  10^{10}$ M$_{\odot}$) for feedback efficiency (or lack of star
  formation). For galaxies with masses lower than this, feedback (or
  star formation suppression) is very efficient while for higher
  masses it is not. The BCT, determined here for the first time,
  should be an important observable with which to confront theoretical
  models of galaxy formation.
\end{abstract}


\section{Introduction}

In the current paradigm for galaxy formation, galaxies form in cold
dark matter halos, which evolve from small, primordial, Gaussian
fluctuations, by gravitational instability.  This mechanism fits well
in the successful LCDM picture which correctly describes the Universe
from the Cosmic Microwave Background at $z=1088$ to local galaxy
clustering \citep{SpergelWMAP03,Percival2dF03}.  One of the strong
predictions of this galaxy formation paradigm is the typical redshift
of dark matter halo formation (i.e.  virialization) as a function of
halo mass and cosmological parameters (e.g., \citet{ST04}).

Given that cosmological parameters have been tightly constrained
(e.g., \cite{SpergelWMAP03}), we can reconstruct the average dark halo
formation history as a function of mass (e.g.,
\citet{PS74,ST99,ST04}).  Naively, the dark matter halo collapse,
i.e. virialization, should trigger baryonic gas transformation into
stars; in addition, subsequent dark matter mergers should produce star
formation episodes. We will show that this simple model is not in
agreement with observations.

All we can observe is the integrated light of galaxies' stellar
population; thus, to compare the theory prediction for the dark
matter halos with observations, the process of how baryons are
transformed into stars needs to be simulated, either through
semi-analytical recipes or by means of hydro-dynamic N-body
simulations. Since no theory of star formation has been yet
established, we do not have a fundamental theory that allows us to
compute from basic principles the star formation efficiency.
Further, complications arise from other phenomena, such as
feedback by stars and AGN that prevent the formation of giant
molecular clouds and therefore reduce star formation efficiency.
Given the complexity of the task of learning about galaxy
formation from numerical simulations, here we take the
complementary approach of placing new observational constraints on
the stellar assembly history as a function of galaxy mass.

In this paper we use about $10^5$ galaxies from the Sloan Digital Sky
Survey Data Release 1 (SDSS DR1) to determine, for the first time, the
amount of baryons that have been transformed into stars as a function
of total stellar mass and time.  This allows us to build the baryonic
conversion tree (BCT), which can then be compared with the dark matter
merger tree. We present such comparison  and show how it can
be used to compute the star formation efficiency and the relative
importance of feedback (or lack of star formation).

Our main findings are:
\begin{enumerate}
\item There is a clear correlation between total stellar mass of
the galaxy and the fraction of gas transformed into stars at $z \geq
1$. The larger the mass the larger the fraction of gas transformed
into stars at high redshift. 

\item If large galaxies need to be formed by $z \sim 1$ in a
  ``monolithic'' fashion, as observations suggest (e.g.,
  \citet{Bower+92,PJDWSSDW98,Lilly+98,BE00,Im+02,Renzini03,Gao+03,Glazebrook+04}),
  high-redshift star formation efficiency needs to be much higher
  (about 10\%) in massive galaxies than in less massive ones.  This
  high star formation activity at early times means the build-up of
  stellar mass does not follow the hierarchical build-up of total
  mass. Stars could form in smaller objects (not in the main
  progenitor only) with lower efficiency, provided that the galaxy
  formation process has some way to synchronise star formation in
  disparate pieces.  The reason being that the above mentioned
  galaxies at $z\sim 1-2$ have stellar populations with almost no
  spread in their stellar ages and derived star formation histories
  consistent with very short times ($< 0.1$ Gyr, see Macarthy et al
  2004).

\item There is an indication that major mergers are not the
principal drivers of star formation.

\item We propose that a threshold for star formation for galaxies with
masses $\sim 10^{10}$ M$_{\odot}$ can explain the findings above. The
existence of such a threshold at low redshift is well documented in
the literature \citep{MK01}, and can be linked to feedback
efficiency. Feedback, i.e. suppression of star formation activity, is
expected to be very inefficient in massive galaxies, while efficient
in less massive galaxies.

\end{enumerate}

\section{Baryon conversion tree from SDSS}

The large size of the spectroscopic sample of the SDSS DR1 provides
the means to analyse statistically properties of galaxies inferred
from their spectra. Further, SDSS spectra have two important
characteristics: a large wavelength coverage ($\sim$ 3500 \AA\, --
8500 \AA\,) and a relatively high signal-to-noise (S/N 6 to 10 per
resolution element of 2\AA\,).  Previous work
\citep{KauffmannSDSS03,PHJ03,Padmanabhan+03,Kauffmann+04} accurately determined the
total stellar mass of SDSS galaxies at the galaxy observed redshift.

The observed spectrum of a galaxy contains (in principle) the ``fossil
record'' \citep{PHJ03} of the galaxy's star formation history, that is
information on the stellar mass as a function of redshift; but no work
has so far been done to determine the stellar mass formation history
(the BCT). Here we attempt for the first time to do this.

One important limitation is imposed by the fact that the old
stellar populations are orders of magnitude dimmer than young
ones, thus they have a sub-dominant effect in the galaxy spectrum.
As a consequence, it is not possible to determine, from the
observed spectrum, the BCT with arbitrary time resolution,
especially at high redshift. On the other hand, we find that one
can accurately determine the BCT for time bins that are
logarithmically spaced in look-back time.

\subsection{Method}
The SDSS DR1 main sample has apparent magnitude limits $15 \leq m_R
\leq 17.77$ and covers a redshift range $0.005 < z < 0.34$, with a
median redshift of 0.1. We place an additional cut on surface
brightness of $\mu_R < 23.0$ to avoid spurious background
contamination (see \citet{Shen+03}), leaving 96,545 galaxies for this
study\footnote{The calibration of the continuum blue wards of 4500
\AA\, has been improved in DR2. We will check the effect of this by
analysing the DR2 sample in a forthcoming paper.}. Full details of the
SDSS are available at {\tt http://www.sdss.org}. The spectra are
top-hat smoothed to 20\AA\, resolution for comparison with the stellar
population models of \citet{JMDPP04} and emission-line regions are
removed
\footnote{i.e., ignored in the likelihood fitting procedure. In
particular the regions excluded are: 3700-3760 \AA\, 4840-5200 \AA\,
and 6500-6800 \AA.}  since these have a complicated dependence on the
geometry of the ionising region and do not carry much information
about the underlying stellar population. The principal strength of
MOPED is that instead of depending on a few, possible contaminated,
lines indices, it uses the whole of the rest of the spectrum in an
optimally-weighted way, which extracts essentially all of the star
formation history information.

\begin{figure}
\includegraphics[width=8.5cm]{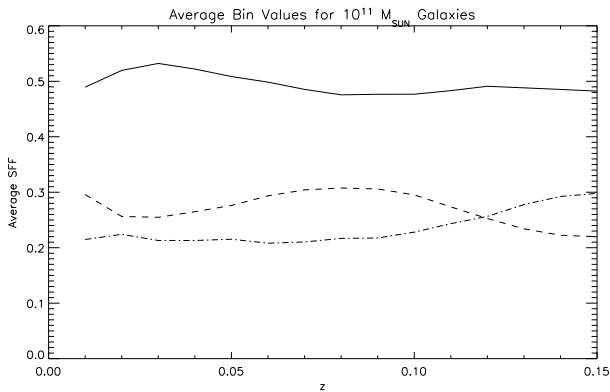}
\caption{Recovered average stellar mass fractions in the oldest three
bins, for galaxies with masses in the range $1-3 \times 10^{11}
M_\odot$.  Notice that there is no significant trend at low redshift,
which might indicate a problem arising from differences between the
small fibre aperture and the larger aperture used for the photometry.}
\label{fig:bctfig}
\end{figure}

We recover the mass of stars created in 10 time bins ($\delta
M_{*}(t_i)$ where $i=1,..,10$), which are equally spaced
logarithmically in look-back time, separated by factors of 2.07.
Table~\ref{table:bins} shows the centre and boundaries of the bins
both in look-back time and redshift. For each time bin we also recover
the average metallicity of the stars ($Z_{*}(t_i)$). The final
parameter is an overall dust parameter for each galaxy at the observed
redshift ($D_{z_o}$); we assume an extinction curve as the one
determined for the LMC \citep{LMC03}. However we are not too sensitive
to the particular extinction curve as we have experimented with a
variety of dust screens and found little variation in the shape of the
recovered star formation.  We use a Salpeter initial mass function,
with a power-law exponent of $X=-1.35$.  To model the galaxy spectra
as a function of these parameters, we use the stellar population
models\footnote{Available at {\tt
http://www.physics.upenn.edu/$\sim$raulj}.} of \citet{JMDPP04}. We
should point out here that, although in principle a maximum likelihood
analysis could recover the star formation history of individual
galaxies, the parameters for a single galaxy are not tightly
constrained. In practice we statistically recover the {\em average}
stellar assembly history from the full DR1.  To do this, galaxies are
weighted inversely with $V_{\rm max}$, the volume in which they could
be found and still satisfy the selection criteria for the survey.  For
this, the evolution of the stellar population and spectrum are
included, but no size evolution is assumed.  Concerns have been raised
that the calibration of the spectra is done by using photometry with a
larger aperture; this issue has been addressed by Glazebrook et
al. (2003), who found that, on average, the colours from the fibres
and from the photometry are consistent (on average), so this should
not be a concern for the sample as a whole.  To support this further,
we show in fig.~\ref{fig:bctfig} the fraction of stellar mass
contributed by the oldest three bins, for galaxies selected to be in a
fixed mass range, as a function of the redshift of the galaxy.
Encouragingly, there is no evidence of any significant trend towards
low-redshift, where the aperture effects might be expected to be a problem.

\begin{figure}
\includegraphics[width=8.5cm]{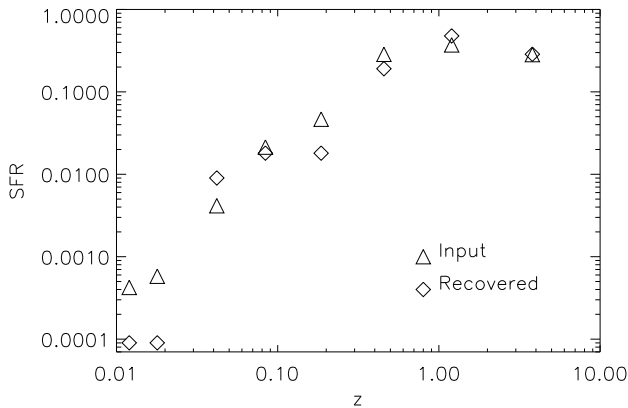}
\caption{Input (diamonds) and average recovered (triangles) SFR for a sample of 500 synthetic
galaxies with wavelength-dependent relative noise properties typical
of the SDSS, a calibration uncertainty (15\% at the blue end,
tapering to zero at 4000 Angstroms), and a variable amount of
line-filling of the $H\delta$ line (mean 0.5 of the model depth, rms 0.2).}
\label{fig:bcttest}
\end{figure}

In addition to this test, we have performed further checks on the
MOPED technique.  In \cite{HPJD04}, we showed (in the supplementary
information on the Nature web-site) that MOPED could recover the star
formation correctly, given an input SFR which matched the SFR we
claimed.  To test this more thoroughly, we have generated synthetic
spectra for 500 galaxies which have a SFR peaking at $z=1$,
corresponding to the broad conclusions of SFR studies
pre-\cite{HPJD04}.  We also include wavelength-dependent relative
noise, characteristic of typical SDSS galaxies, and a systematic
calibration offset at the blue end.  We also allow the H$\delta$ line
to be randomly partially filled with emission.  This should be a
rather thorough test that the results are not biased by noise,
calibration or line-filling.  The input and average recovered star
formation rates are shown in Fig.~\ref{fig:bcttest} which illustrates that
indeed MOPED does recover the input star formation history without any
biases.

\begin{figure}
\includegraphics[width=8.5cm]{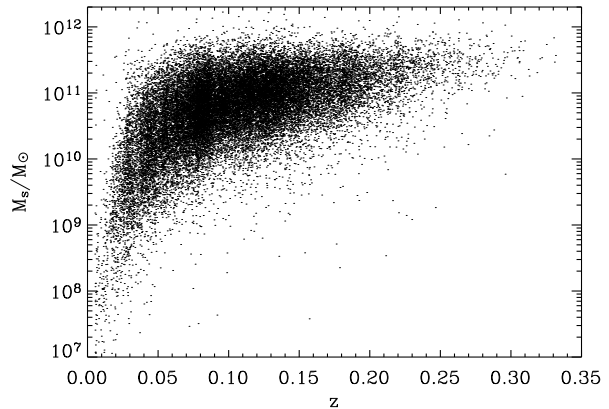}
\caption{Recovered total mass in stars as a function of redshift for
the whole sample.}
\label{fig:bctfig0}
\end{figure}

Although we are mostly interested in the 10 $\delta M_{*}(t_i)$,
in total we have 21 parameters which we want to constrain from
96,545 SDSS spectra (each of which has about $2000$ elements)
using a maximum likelihood approach. This is an extremely
computationally expensive task, which we can handle by resorting
to 1) a data compression algorithm and 2) the now widely used
Markov Chain Monte Carlo (MCMC) technique.

The data compression algorithm MOPED \citep*{HJL00} enables us to
explore efficiently the parameter space and place error bars on the
recovered parameters.  The algorithm linearly combines the 2000 flux
elements in each spectrum in 21 MOPED coefficients, one for each
parameter, that contain all the relevant information (i.e. the method
is lossless). For further details about MOPED see \citet*{HJL00}. We
have already demonstrated the usefulness of MOPED and the MCMC
implementation to recover galaxy physical properties from different
galaxy spectroscopic samples, including the SDSS EDR and DR1
\citep*{RJH01,PHJ03,HPJD04}.  Essentially, it works by weighting the
flux data in a way which preserves the information inherent in the
complete spectrum, so basically no information is lost and the
procedure is almost equivalent to using all of the flux values in the
spectrum.  For an individual galaxy, there may still be residual
degeneracies in the solution, but we have demonstrated with tests that
for a very large sample, such as in SDSS, the average solution
recovers the input extremely well.

\begin{figure}
\includegraphics[width=8.5cm]{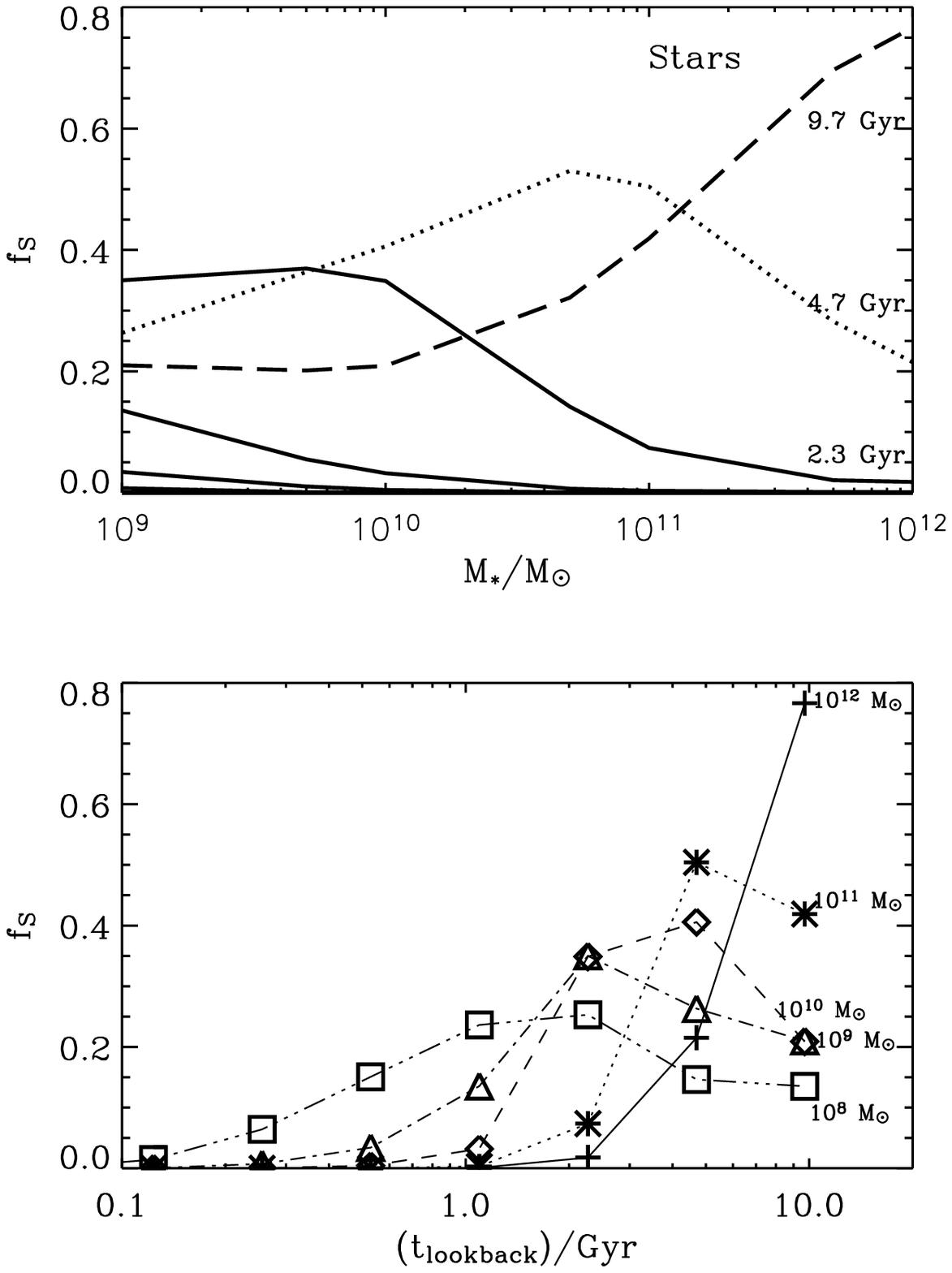}
\caption{Top panel: fraction of stellar mass of the galaxy formed in
different time bins as a
function of total stellar mass. Dashed
line corresponds to the oldest bin (9.7 Gyr), dotted line to the
second oldest (4.7 Gyr) and solid line to the third oldest (2.3
Gyr). The other solid lines correspond to younger time bins (1.1,
0.53, 0.26). Bottom panel: fraction of the observed total stellar mass of
the galaxy created as a function of time for different galaxy stellar masses
(determined at the observed redshift).} \label{fig:bctfig1}
\end{figure}

\begin{table*}
\caption{Centre and boundaries of the bins in redshift ($z$) and
look-back time $t_{\rm lookback}$ in Gyr} \label{table:bins}
\begin{tabular}{l||c|c|c||c|c|c}
\hline
bin \# & lower $z$&centre $z$&upper $z$&lower $t_{\rm lookback}$(Gyr)&centre $t_{\rm lookback}$(Gyr)&upper $t_{\rm lookback}$(Gyr) \\
\hline
1   & 0.0007  &  0.001 & 0.00145 &  0.00966        & 0.014   &  0.0200    \\
2   & 0.00145 & 0.0021 & 0.003   &  0.0200         & 0.029   &  0.0414     \\
3   & 0.003   &  0.006 & 0.0063  &  0.0414         & 0.06    &  0.0857      \\
4   & 0.0063  &  0.012 & 0.013   &  0.0857         & 0.12    &  0.1776      \\
5   & 0.013    & 0.0179 & 0.027   &  0.1776         & 0.26    &  0.3677      \\
6   & 0.027    & 0.0419 & 0.057   &  0.3677         & 0.53    &  0.7614      \\
7   & 0.057    & 0.0839 & 0.125   &  0.7614         & 1.10    &  1.5767      \\
8   & 0.125   & 0.186  & 0.287   &  1.5767         & 2.27    &  3.2650     \\
9   & 0.287   & 0.456  & 0.786   &  3.2650         & 4.70    &  6.7609      \\
10  & 0.786   & 1.755  & 5.000   &  6.7609         & 9.7     &  12.000      \\
\hline
\end{tabular}
\end{table*}

The SDSS survey is a magnitude-limited survey and here we aim at
deriving {\em average} properties, i.e. the stellar assembly history,
of galaxies as a function of their present-day stellar mass.  In
principle, this could introduce a bias since for a given mass,
galaxies with young stellar population, tend to be brighter than
galaxies with an old stellar population; thus blue galaxies which form
stars recently, would be preferred over passively evolving
ones. However this does not matter for our purposes for the following
reasons: {\it a)} galaxies have been effectively selected by mass from
a complete redshift and magnitude-limited sample. This can be seen in
Fig.~\ref{fig:bctfig0} where the recovered total mass is plotted as a
function of the observed redshift. Note that as expected, for a
mass-selected sample, less massive galaxies are at low redshifts while
most massive galaxies occupy the whole redshift range; {\it b)} we
will determine galaxy properties averaged in mass-bins; {\it c)} the
recovered mass fractions for each time bin for each galaxy are
properly shifted according to the observed redshift of the galaxy and
assuming that the size of galaxies does not evolve over the period of
time covered by the observed redshift range.  Note that the results on
stellar mass fractions are insensitive to the redshift of the galaxies
used - see Fig.~\ref{fig:bctfig0}. To determine average
properties as a function of mass we average our results in mass with
mass bins equally-spaced logarithmically by factors of ten.

\subsection{Results}

We find that more massive galaxies transform their gas into
stars {\em earlier} than less massive ones.  The top panel of
Fig.~\ref{fig:bctfig1} shows for each time bin how many baryons
(as a fraction of the observed total stellar mass) were converted
into stars, $f_s$, as a function of observed stellar mass
$M_{*}=M_*(t=0)$. The different lines correspond to different time
bins: dash oldest, dots second-oldest and continuous lines denote
younger bins. The bottom panel shows the stellar mass assembly
history $\delta M_{*}(t_i)$ for different observed stellar
masses.

There is a clear correlation between baryon conversion efficiency
and present-day stellar mass. This can be see from
Fig.~\ref{fig:bctfig1}.  In galaxies with $M_{*}$ larger than
$3 \times 10^{11}$ M$_{\odot}$, more than 60\% of their present
stellar mass was already in place at redshift $1.7$. In terms of
look-back time it means that the stars of massive galaxies
($M_{*}\sim 10^{12}$ M$_{\odot}$) were essentially formed (if
not in place) more than 9 Gyr ago, or just $\sim 4$ Gyr after the
big bang. Conversely, for $M_{*} < 10^{9}$ M$_{\odot}$,
more than half of their stellar mass remains unformed at $z=0.2$ or $\sim$ 3
Gyr ago.

\begin{figure}
\includegraphics[width=8.5cm]{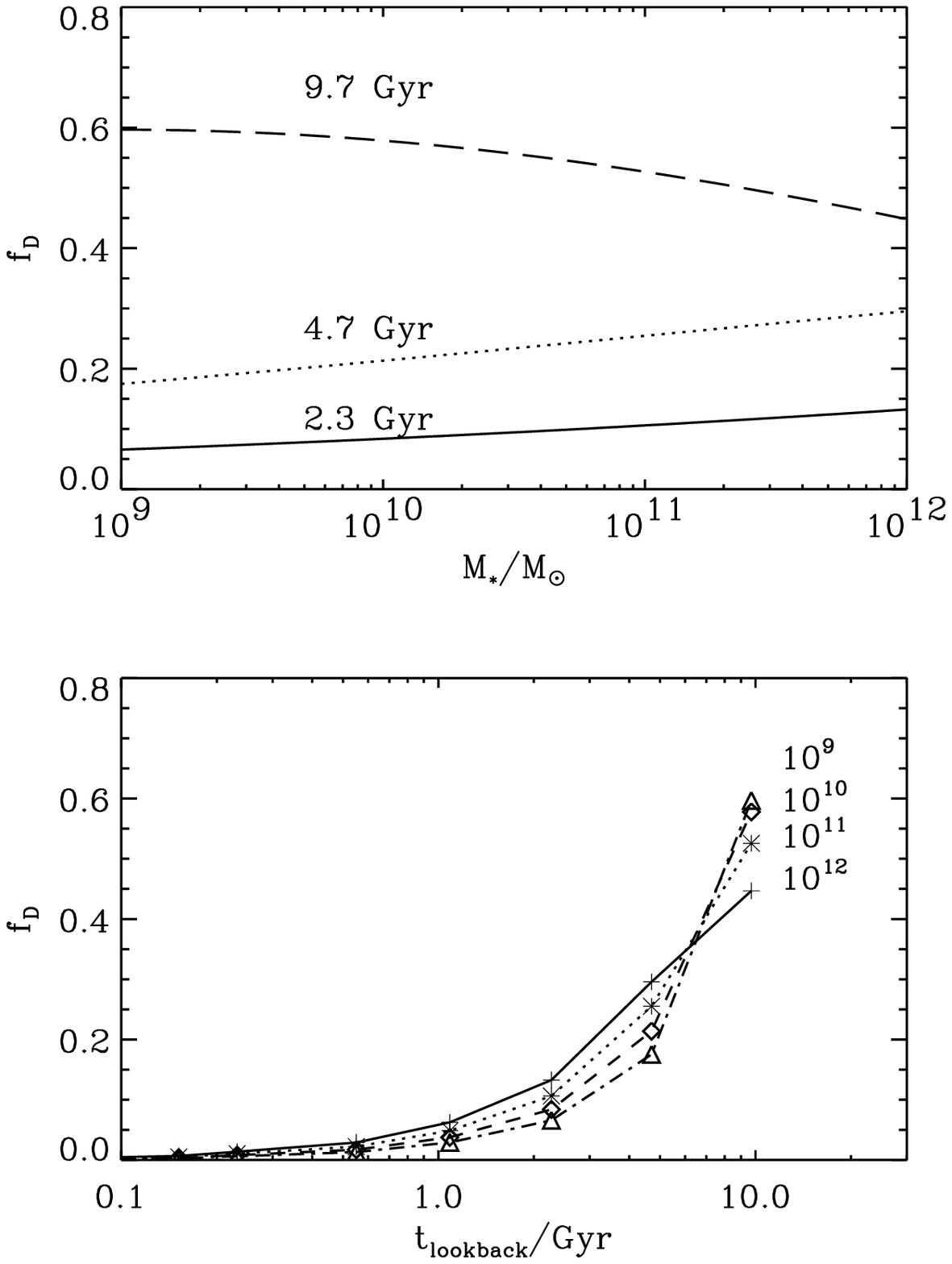}
\caption{Top panel: fraction of the final dark matter mass that
has been virialised in the most massive progenitor, as a function of the total stellar mass
of the galaxy (at the observed redshift). The different lines
correspond to the amount of dark matter that is virialised in the
given time bin. Bottom panel: fraction of the final dark matter
mass that has been virialised in the most massive progenitor  as a function of $t_{\rm lookback}$.
The different lines correspond to different total stellar masses
of the galaxy at the observed redshift.} \label{fig:bctfig2}
\end{figure}

\section{Dark matter and baryon assembly history}

We now compare the stellar formation history we just recovered,
with the dark matter assembly history. To compute the mass of dark
matter halos as a function of time we use two approaches: first,
we generate multiple realizations ($10^4$) of the merger history
of dark matter halos of several masses in the range $10^9$ to
$10^{13}$ M$_{\odot}$. This is done using the algorithm described
in \citet{SK99} for the standard LCDM cosmology ($\Omega_0=0.27$,
$\Lambda_0=0.73$, $h=0.71$, \citet{SpergelWMAP03}). This algorithm
reproduces the merger histories of halos seen in N-body
simulations of structure formation \citep{SLKD00,WBPKD02},
especially at low redshift, which is the range of interest. One
free parameter in the algorithm is the value of the mass that is
considered accreted instead of merged.  Agreement with CDM
simulations is achieved when everything below 1\% of the final
halo mass is considered accreted and this is the value we use.
Second, we use the fitting formula for the mass accretion history
from \citet{WBPKD02}. This is obtained from numerical N-body
simulations performed with the ART code \citep{Kravtsov+97}. We
find that both approaches show the same qualitative behaviour
illustrated in Fig.~\ref{fig:bctfig2}, however since the Extended
Press-Schechter approach, at the base of the \citet{SK99}
algorithm, is not a perfect fit to N-body simulations, especially
at high-redshift and for very massive halos, we will present here
only results obtained using the second approach.

We compute the dark matter mass of the most massive progenitor that is
virialised in each redshift bin as a function of the total mass of the
dark halo. The top panel of Fig.~\ref{fig:bctfig2} shows the fraction
of the final dark matter mass that has been virialised in the most
massive progenitor, $f_D$, in a given redshift bin (line styles same
as in Fig.~\ref{fig:bctfig1}) as a function of the final stellar mass
in the dark halo.  The stellar mass of the dark halo is obtained from
the dark one assuming the universal baryon fraction $f_b$ as
determined by WMAP ($\Omega_{\rm CDM}/\Omega_b=4.8 \equiv f_{\rm
DM/b}$; \citet{SpergelWMAP03}), and that at $z\sim 0$ only about $6$\% of
the baryons are in stars \citep{Fukugita2003}.

The bottom panel of Fig.~\ref{fig:bctfig2} shows the mass
assembly history of the dark halo.

A comparison of Fig.~\ref{fig:bctfig1} and Fig.~\ref{fig:bctfig2}
indicates that dark matter assembly and formation of stars do not
follow each other. For example, for $M_{*}> 10^{12}$ M$_{\odot}$ less
than 50\% of the dark matter is assembled in the main progenitor at
$z>1.7$ ($t_{\rm lookback} \sim 9.7$ Gyr), while more than $75$\% of
the stellar mass is already formed. On the other hand, for stellar
masses smaller than $10^{11}$ M$_{\odot}$, $20$\% of the stellar mass
is formed in the same time bin while already $60$\% of the dark matter
is in place. This hints at a role of early stellar feedback in these
halos as we discuss later in \S 4.

While in the hierarchical LCDM model for structure formation the
more massive CDM structures form late, we find observational
evidence for early star formation of giant galaxies. This can
happen because of two reasons: {\em a)} massive galaxies are
formed by mergers with smaller ones, each carrying an evolved
stellar population {\em b)} these massive galaxies are already in
place at high-redshift and the dark matter halo has been assembled
at the same time as the stellar population.  If the dark matter
halo collapse triggers star formation then one would expect {\em
a)} to be the case; however observations of e.g., old elliptical
galaxies at high redshift (e.g. \citet{D+96,NDJH03,Daddi+04,Saracco+03}) seem
to support the second scenario, at least in some cases.

\subsection{The number of progenitors of galaxies}

It is clear that since massive dark halos are predicted, {\it on
average}, to assemble later than the stellar population they contain,
the stars may naturally have formed previously in smaller dark halos
that subsequently merged.  By considering the (dark) mass accretion
history of the most massive progenitor of a halo, and the
star-formation rate, we can constrain the minimum number of
progenitors forming a halo, and the star formation efficiency as
outlined below. In other words, if the most massive progenitor at a
given redshift $z$ carries enough baryons to form all the stars that
should be in place by $z$ then only one progenitor (the most massive
one) is needed.  If the minimum number of progenitors forming a halo
is larger than 1, massive galaxies must have form by mergers with
smaller ones, each carrying an evolved stellar population. However, if
the minimum number of progenitors is 1, then the LCDM paradigm can
accommodate old elliptical galaxies at high redshift.

For example, consider a galaxy with stellar mass $M_{*}=10^{12}$
M$_{\odot}$ (see Fig.~\ref{fig:bctfig1} and Fig.~\ref{fig:bctfig2}).
In the oldest time bin 76\% of the stellar mass is already in place
($M_{*}(t_{10})=0.76 M_{*}$), yet the virialised fraction of the dark
matter halo is less than 50\%. If we assume case a) (that is star
formation happens uniformly in all halos and sub-halos that ultimately
end up forming the final galaxy, and that 100\% of the dark matter is
virialised at all times) then the fraction of the total baryonic mass
converted into stars was 4.6\%. 
On the other hand, observations of old elliptical galaxies
at high redshift can be explained within the LCDM paradigm if we
assume that the main virialised progenitor harboured all the stars
observed, in this case the fraction of mass converted to stars in the
dark matter halo must have been $\sim$ 10\%. This requires only a
modest enhancement in star formation efficiency at high redshift, but
close to the maximum efficiency observed in giant molecular clouds
today, which is $< 10$\% (e.g., \citet{PN02}).

More specifically, we can assume that the stellar mass $M_{*}(t_{10})$,
was in progenitors (the most massive progenitor and possibly other sub
halos), whose cumulative dark matter must have been at least
$M_{*}(t_{10})\times f_{\rm DM/b}/f$, where $f\le 1$ parameterises the
star formation efficiency, and the fraction of baryons that gets
turned into stars, and depends on the mass of the sub-halos. We
approximate $f(M_{\rm dark})$ as $f_s(M_{\rm dark}/f_{\rm DM/b} \times
0.1)$ (as 0.1 is the minimum efficiency in the oldest time bin).

The minimal number of progenitors can thus be obtained by minimisation
as a function of the sub halo dark mass $M_{SH}$:
\begin{equation}
N(M_{\rm SH})=1+{\rm max}\left\{\frac{f_{\rm DM/b} M_{*}(t_{10})
/f(M_{\rm SH})-M_{\rm dark}(t_{10})}{M_{\rm SH}},0\right\}
\end{equation}
We obtain a minimum number of $1$ and $f \sim 10$\%, in agreement with
the more heuristic argument above. If the main progenitor did not have
enough baryons to accommodate $M_{*}(t_{10})$, then the minimum number
of progenitors would be greater than one. If the minimum number of
progenitors is one and the star formation efficiency is constrained to
be reasonable, then all the old stellar population could have been
formed in the main progenitor.

Thus this suggests that in the LCDM paradigm, the massive old galaxies
that we see today could have been made of mergers of few progenitors,
each carrying an old stellar population, in agreement with other
indications that elliptical galaxies are already formed at $z>1$
(e.g. \citet{Bower+92,PJDWSSDW98,Lilly+98,BE00,Im+02,Renzini03,Gao+03,Saracco+03,Glazebrook+04}).
A small number of mergers can also naturally explain the tightness of
the observed colour-magnitude relation \citep{Bower+92}.

\section{Time evolution of star formation efficiency}

\begin{figure}
\includegraphics[width=8.5cm]{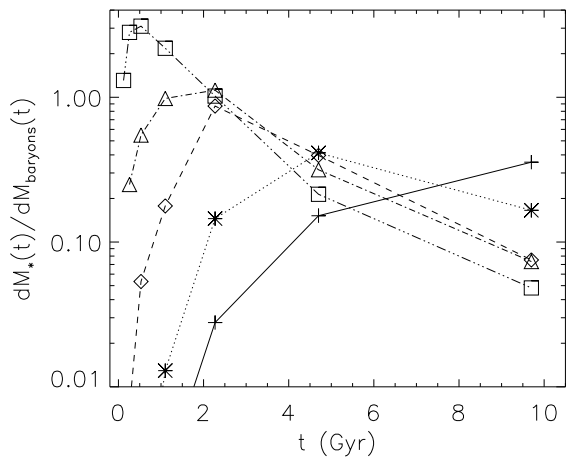}
\caption{Ratio of mass transformed into stars in a galaxy to the
baryons that are accreted onto the
main progenitor, which typically contains $>$ half the mass.  Crosses correspond to a stellar
mass of $10^{12}$, asterisks to $10^{11}$, diamonds to $10^{10}$,
triangles to $10^9$ and squares to $10^8$ M$_{\odot}$. A value for
$dM_*(t)/dM_{\rm baryons}(t)$ smaller than 1 means that not all
the baryons (the nucleosynthesis value as recently constrained by
WMAP) have been transformed into stars, while a value larger than
1 indicates that gas previously available in the galaxy has been
turned into stars:  either the recent accretion has triggered star
formation in the main progenitor, or it is going on elsewhere. 
Note that for stellar masses below $10^{10}$
M$_{\odot}$, the star formation efficiency peaks at $t_{\rm
lookback}\sim 1-2$ Gyr.} \label{fig:bctfig3}
\end{figure}

For each of the time bins we compute the ratio of the newly-formed
stellar mass to the baryonic mass added to the main progenitor,
assuming the nucleosynthesis baryon fraction. Fig.~\ref{fig:bctfig3}
shows the above ratio as a function of look-back time ($t_{\rm
lookback}$) for different masses: crosses correspond to a stellar mass
of $10^{12}$ M$_{\odot}$, asterisks to $10^{11}$ M$_{\odot}$, diamonds
to $10^{10}$ M$_{\odot}$, triangles to $10^{9}$ M$_{\odot}$ and
squares to $10^8$ M$_{\odot}$. This is a measurement of how much gas
is transformed into stars as a function of the newly-added baryons. A
value of $1$ indicates that the mass of baryons accreted to the main
progenitor matches the mass converted into stars. A value higher than
$1$ shows that the accretion or merger was accompanied by a greater
mass of triggered star formation somewhere in the galaxy.
Fig.~\ref{fig:bctfig3} clearly shows that for stellar masses above
$10^{10}$ M$_{\odot}$ this ratio is never greater than $1$.  Clearly
we are comparing mass accreted on to the main progenitor with stars
created in any of the progenitors, and this should be borne in mind in
interpreting Fig.~\ref{fig:bctfig3}.  However, since the main
progenitor contains $\sim 50\%$ of the final mass even at a lookback
time of 10 Gyr, and this fraction is weakly dependent on mass, the
efficiency of conversion of baryons to stars in the galaxy as a whole
is unlikely to alter the main conclusions of Fig.~\ref{fig:bctfig3},
where the differences between objects of different present-day masses
typically far exceed a factor of 2.

For the most massive galaxies at early times this measure of star formation
efficiency is close to 40\%. For stellar masses below
$10^{10}$M$_{\odot}$, the efficiency is only of about 6-8\% at $t_{\rm
lookback}=10$ Gyr, but grows to 100\% at $t_{\rm lookback}=2$ Gyr, and
then decreases.

For galaxies with stellar mass smaller than $10^{9}$ M$_{\odot}$, this
increase in star formation efficiency rises until $t_{\rm
lookback}\sim 0.5$ Gyr, at which point it reaches 300\% efficiency,
which means that more gas is transformed into stars than the baryons
brought into the parent dark halo by accretion.  This points toward a
picture where these low-mass gas-rich galaxies see a lot of their gas
reservoir transformed into stars due, for example, to a merger or
accretion event.  Another possibility is that star formation is
proceeding rapidly in other sub-halos, which subsequently merge with
the parent. This scenario is not supported by the merger histories of
low mass halos (e.g. Sheth \& Tormen 2004; Wechsler et al. 2002) since
small halos at present time have almost no merging from smaller
subhalos.

Thus massive galaxies have a high star-formation efficiency at early
times and then evolve ``passively'', with fresh infall of gas being
suppressed or turned into stars with low efficiency, possibly because
it is likely to be too hot. Small galaxies seem to accrete mass
passively at early times and form stars very efficiently later.

Conversely, the probability distribution for dark halo merger
events peaks at higher redshift for small halos and at lower
redshift for large halos.  In a $\Lambda$-dominated Universe,
merger probability is suppressed at $z<1$ especially in low
density regions, where small galaxies are most likely to be.  Thus
there seems to be no correlation between halo virialization or
dark matter merger events and star formation efficiency. 

However, one could imagine explaining this trend, for example, by
postulating the existence of a ``threshold'' for star formation:
once this threshold is crossed, all available baryons are turned into
stars (as in an ``infall model''), then afterward galaxies evolve
passively (as in a ``closed box model''). In this toy model, if this
threshold is crossed at very early times in the progenitors of massive
galaxies, one would expect these galaxies to form stars very
efficiently early on then evolve passively. For progressively smaller
galaxies this threshold is met at increasingly later times.

An alternative explanation is that stellar feedback is the responsible
for the lack of star formation in small galaxies at early times. Since
the escape velocity in these systems is smaller than in more massive
galaxies, gas can leave the dark matter halo more easily. Only the
very massive systems are able to retain their gas and convert a
majority of their gas into stars. If so we find that $M_* \sim
10^{10}$ M$_{\odot}$ is the characteristic mass that defines the
border between efficient and inefficient feedback.

\section{Conclusions}

We have determined for the first time the baryonic conversion tree
for galaxies.  We were able to do this with observations at
$z<0.3$, using 96545 galaxies from the SDSS DR1 spectroscopic
sample.

In the hierarchical structure formation model, massive {\em dark
halos} form (i.e. virialize) later than smaller halos, from mergers of
smaller units (e.g. \citet{lc93,lc94,ljl03}).  This model has been
thoroughly tested against numerical cold dark matter simulations.
Naively one could expect that the dark matter halo collapse should
trigger baryonic gas transformation into stars; in addition subsequent
dark matter mergers should produce star formation episodes.

Instead, for the the stellar assembly history we find that: the more
massive galaxies have old stellar population and massive, old
elliptical galaxies are already in place at $z \sim 1$.  This has been
known for a long time and sometimes it is referred to as
``down-sizing'', \citep{CSC96}.  We find that massive galaxies
have transformed more gas into stars at higher redshift (in agreement
with high z observations e.g., \cite{kodama04}), and then star
formation was suppressed, while less massive galaxies transform more
gas into stars at low redshift.

So one should not be surprised to see abundant red objects at high
redshift in the LCDM paradigm, these objects can form in virialised
halos if star formation efficiency is high.  Indeed, \cite{JFDTPN99}
have shown that single-halo hydro-dynamical models would require an
increased star formation efficiency for more massive galaxies, higher
than the few percent found today in giant molecular clouds, in
agreement with the value determined
from the ``fossil record''  in the present work.
Our findings, based only on observations at $z < 0.35$ (the ``fossil record''), are in agreement with a suite of independent, $z>1$, observations:
observations of old elliptical galaxies at high redshift
\citep{D+96,Spinrad+97,NDJH03}, indications that elliptical galaxies
are already formed at $z>1$
(e.g. \citet{Bower+92,PJDWSSDW98,Lilly+98,BE00,Im+02,Renzini03,Gao+03,Glazebrook+04}). It
also nicely explains the tightness of the colour-magnitude relation
\citep{Bower+92}.

On the other hand, we find that small galaxies seem to accrete mass
passively at early times and see a lot of their gas reservoir
transformed into stars at late times.  Since the probability
distribution for dark halo mergers peaks at low redshift for massive
halos and high redshift for small halos, we conclude that dark matter
mergers and star formation are not correlated.  We speculate that one
possibility to explain the apparent and illusory anti-hierarchical
nature of the stellar assembly history is the existence of a threshold
for star formation: once the threshold is crossed all available
baryons are turned into stars (``infall model'') and afterward
galaxies approximatively evolve passively. The threshold is met at
very early time for massive galaxies and a later time for less massive
ones (see e.g. \citet{HJ99}).

A star formation threshold has been observed in disk galaxies by
\citet{MK01} and there has been some recent additional evidence from
the formation of dust lanes in disk galaxies \citep{DYB04} that this
threshold may take place at $V_c \sim 100$ km s$^{-1}$, in agreement
with the findings of \citet{VOJ02} and \citet{kannappan04} who also
found a transition at about $100$ km s$^{-1}$ for star formation
efficiency. This $V_c$ value corresponds to the characteristic mass
found here ($M_* \sim 10^{10}$ M$_{\odot}$), that defines the border
between efficient and inefficient star formation.  This characteristic
mass has been related to feedback efficiency threshold (e.g., Dekel \&
Silk 1986, Dekel \& Woo 2003 and references therein).

As we do not yet have a fundamental theory for galaxy formation and given the
complexity involved in studying the process with hydrodynamic-N body
simulations, we hope that this new determination of the baryonic conversion
history will be a useful observable to gauge galaxy formation models against.\\

\section*{acknowledgments}
LV and RJ thank the ``Centro de Investigationes de Astronomia''
(CIDA), where part of this work was carried out, for hospitality. LV
and RJ thank Sheila Kannappan for stimulating discussions and the
anonymous referee for insightful comments. The work of RJ is partially
supported by NSF grant AST-0206031.


\end{document}